\overfullrule=0pt
\input harvmac
\lref\BerkovitsYR{
  N.~Berkovits and O.~Chandia,
  ``Superstring vertex operators in an AdS(5) x S**5 background,''
Nucl.\ Phys.\ B {\bf 596}, 185 (2001).
[hep-th/0009168].
}

\lref\BerkovitsRB{
  N.~Berkovits,
``Covariant quantization of the superparticle using pure spinors,''
JHEP {\bf 0109}, 016 (2001).
[hep-th/0105050].
}

\lref\GalperinAV{
  A.~Galperin, E.~Ivanov, S.~Kalitsyn, V.~Ogievetsky and E.~Sokatchev,
``Unconstrained N=2 Matter, Yang-Mills and Supergravity Theories in Harmonic Superspace,''
Class.\ Quant.\ Grav.\  {\bf 1}, 469 (1984).
}

\lref\vallilo{
  B.~Vallilo and L.~Mazzucato,
  ``The Konishi Multilpet at Strong Coupling,''
JHEP {\bf 1112}, 029 (2011).
[arXiv:1102.1219 [hep-th]].
}

\lref\MikhailovAF{
  A.~Mikhailov,
 ``Finite dimensional vertex,''
JHEP {\bf 1112}, 005 (2011).
[arXiv:1105.2231 [hep-th]].
}


\lref\mikh{
  A.~Mikhailov and R.~Xu,
  ``BRST cohomology of the sum of two pure spinors,''
[arXiv:1301.3353 [hep-th]].
}

\lref\minahan{
  J.~Minahan,
  ``Holographic three-point functions for short operators,''
JHEP {\bf 1207}, 187 (2012).
[arXiv:1206.3129 [hep-th]].
}



\lref\BerkovitsBT{
  N.~Berkovits,
  ``Pure spinor formalism as an N=2 topological string,''
JHEP {\bf 0510}, 089 (2005).
[hep-th/0509120].
}



\lref\BerkovitsFE{
  N.~Berkovits,
  ``Super Poincare covariant quantization of the superstring,''
JHEP {\bf 0004}, 018 (2000).
[hep-th/0001035].
}


\lref\MazzucatoJT{
  L.~Mazzucato,
  ``Superstrings in AdS,''
[arXiv:1104.2604 [hep-th]].
}


\lref\HeslopNP{
  P.~Heslop and P.~S.~Howe,
  ``Chiral superfields in IIB supergravity,''
Phys.\ Lett.\ B {\bf 502}, 259 (2001).
[hep-th/0008047].
}







\lref\SohniusWK{
  M.~F.~Sohnius,
  ``Bianchi Identities for Supersymmetric Gauge Theories,''
Nucl.\ Phys.\ B {\bf 136}, 461 (1978).
}

\lref\ArutyunovGA{
  G.~Arutyunov and S.~Frolov,
  ``Foundations of the $AdS_5 \, x \, S^5$ Superstring. Part I,''
J.\ Phys.\ A {\bf 42}, 254003 (2009).
[arXiv:0901.4937 [hep-th]].
}

\lref\MetsaevIT{
  R.~R.~Metsaev and A.~A.~Tseytlin,
  ``Type IIB superstring action in AdS(5) x S**5 background,''
Nucl.\ Phys.\ B {\bf 533}, 109 (1998).
[hep-th/9805028].
}



\lref\HoweSRA{
  P.~S.~Howe and P.~C.~West,
  ``The Complete N=2, D=10 Supergravity,''
Nucl.\ Phys.\ B {\bf 238}, 181 (1984).
}

\def\bar{\overline}
\def\ads{AdS_5\times S^5}
\def\a{{\alpha}}
\def\ve{{\varepsilon}}

\def\ad{{\dot a}}

\def\lh{{\widehat \lambda}}

\def\wh{{\widehat w}}
\def\bd{{\dot \b}}
\def\l{{\lambda}}
\def\lb{{\overline\lambda}}
\def\wb{{\overline w}}
\def\lhb{\widehat{\overline\lambda}}
\def\whb{\widehat{\overline w}}

\def\cN{{{\cal N}=4}}
\def\lb{{\overline\lambda}}
\def\b{{\beta}}

\def\g{{\gamma}}

\def\d{{\delta}}
\def\e{{\epsilon}}
\def\s{{\sigma}}

\def\O{{\Omega}}
\def\half{{1\over 2}}
\def\p{{\partial}}
\def\ads{{{$AdS_5\times S^5$}}}

\def\t{{\theta}}

\def\ad{{\dot\a}}
\def\bd{{\dot\b}}

\def\S{{\Sigma}}

\def\tb{{\bar\theta}}

\def\lb{{\bar{\lambda}}}
\def\Dt{{\widetilde{D}}}

\def\DJJ{{\tilde{D}^{J'}_{J}}}
\def\lti{{\widetilde{\l}}}
\def\aba{{\bar\a}}
\def\bba{{\bar\b}}

\def\wwb{{\bar{v}}}
\def\www{{v}}
\def\psih{\hat{\psi}}

\Title{\vbox{\baselineskip12pt
\hbox{ICTP-SAIFR/2012-006}}}
{{\vbox{\centerline{Harmonic Superspace from the 
 }
\smallskip
\centerline{$AdS_5\times S^5$ Pure Spinor Formalism}}} }
\bigskip\centerline{Nathan Berkovits\foot{e-mail: nberkovi@ift.unesp.br} and
Thiago Fleury\foot{e-mail: tfleury@ift.unesp.br}}
\bigskip
\centerline{\it ICTP South American Institute for Fundamental Research}
\centerline{\it Instituto de F\'\i sica Te\'orica, UNESP - Univ. 
Estadual Paulista }
\centerline{\it Rua Dr. Bento T. Ferraz 271, 01140-070, S\~ao Paulo, SP, Brasil}
\bigskip

\vskip .3in

On-shell supergravity vertex operators in an $AdS_5\times S^5$ background
are described in the pure spinor formalism by the zero mode cohomology of a
BRST operator. After expanding the pure spinor
BRST operator in terms of the $AdS_5$
radius variable, this cohomology is computed using ${\cal N}=4$
harmonic superspace variables and explicit superfield expressions are
obtained for the behavior of supergravity vertex operators near the boundary
of $AdS_5$.

\vskip .3in

\Date {December 2012}

\newsec{Introduction}

Although the superstring 
worldsheet action in an $AdS_5\times S^5$
background is well-known both in the Green-Schwarz \MetsaevIT\ \ArutyunovGA\  
and pure spinor \BerkovitsFE\ \MazzucatoJT\ 
formalisms, an explicit superfield construction of
$AdS_5\times S^5$ supergravity vertex
operators is still an open problem. Superfield expressions for the
dual half-BPS super-Yang-Mills gauge-invariant operators have been constructed
using d=4 {\cal N}=4 harmonic superspace \GalperinAV\ \ref\howewestn{
P.S. Howe and P.C. West, ``Nonperturbative Green's functions in theories 
with extended superconformal symmetry,'' Int. J. Mod. Phys. A{\bf 14}  
2659 (1999). [hep-th/9509140].}
\ref\ferr{L. Andrianopoli and S. 
Ferrara, ``K-K excitations on AdS(5) x S**5 as N = 4 `primary' superfields, '' 
Phys. Lett. B {\bf 430} 248  (1998). [hep-th/9803171].}, 
however, analogous superfield
expressions for supergravity have only been constructed for the 
field strengths \HeslopNP\ \HoweSRA\ and not for the supergravity gauge fields that 
appear in superstring vertex operators.

In the Green-Schwarz formalism, supergravity vertex operators must preserve
kappa-symmetry and, in the pure spinor formalism, supergravity vertex
operators must preserve BRST invariance. Only the pure spinor formalism
will be discussed here, however, it should be possible to extend our
results for the Green-Schwarz supergravity vertex operators. In any
consistent curved background, Type IIB 
supergravity vertex operators in the pure
spinor formalism are defined by \BerkovitsYR 
\eqn\purever{V = \l^\a \lh^\b A_{\a\b}(x,\t,\widehat\t)}
where $\l^\a$ and $\lh^\b$ are left and right-moving pure spinors and
$A_{\a\b}$ is an ${\cal N}=2$ d=10 bispinor superfield.
The BRST operator is $Q=\l^\a \nabla_\a + \lh^\b \widehat\nabla_\b$
where $\nabla_\a$ and $\widehat \nabla_\b$ are the covariant
${\cal N}=2$ d=10 superspace derivatives.

In this paper, we will compute the BRST cohomology by first
expanding $Q$ near the boundary of $AdS_5$ as
\eqn\expq{Q =  Q_{-\half} + Q_{\half} +
Q_{3\over 2} + ...}
where $Q_n$ is proportional to $z^n$ and
$z$ is the distance from the boundary of $AdS_5$.
In performing
this expansion, it will be convenient to use a ${{PSU(2,2|4)}\over 
{SO(4,1)\times SO(6)}}\times {{SO(6)}\over {SO(5)}}$ supercoset description
of $AdS_5\times S^5$ instead of the usual 
${{PSU(2,2|4)}\over {SO(4,1)\times SO(5)}}$ supercoset description.
We will then argue, making some assumptions, that the BRST cohomology is completely determined by
the cohomology of the first two terms $
Q_{-\half} + Q_{\half}$ and, as expected from
holography, BRST-invariant vertex operators are determined by their behavior
near the boundary of $AdS_5$. Finally, we will compute the zero mode
cohomology of 
$Q_{-\half} + Q_{\half}$ and express the result
in ${\cal N}=4$ d=4 harmonic superspace. In this way, we will obtain
explicit superfield expressions for the behavior of supergravity vertex
operators near the boundary of $AdS_5$.

In section 2 of this paper, we describe the $AdS_5\times S^5$
pure spinor formalism using the 
${{PSU(2,2|4)}\over 
{SO(4,1)\times SO(6)}}\times {{SO(6)}\over {SO(5)}}$ supercoset 
instead of the usual 
${{PSU(2,2|4)}\over {SO(4,1)\times SO(5)}}$ supercoset.
In section 3, we expand the BRST operator as 
$Q = Q_{-\half} + Q_{\half} + Q_{3\over 2} + ...$
where $Q_n$ is proportional to $z^n$ and $z$ is the distance from the $AdS_5$ boundary, and argue
that the cohomology is determined by the first two terms
$Q_{-\half} + Q_{\half}$. 
In section 4, we
restrict to the zero mode cohomology corresponding to supergravity
states and explicitly compute the cohomology of
$Q_{-\half} + Q_{\half}$, thereby obtaining
explicit superfield expressions in ${\cal N}=4$
harmonic superspace for the behavior of supergravity vertex operators
near the boundary of $AdS_5$.
In section 5, we summarize our results and discuss possible applications
such as computation of the massive spectrum and
tree-level scattering amplitudes in $AdS_5\times S^5$.

\newsec{Pure Spinor Formalism with 
${{PSU(2,2|4)}\over{SO(4,1)\times SO(6)}}$ Supercoset}

In this section, the pure spinor formalism in an \ads\ background
will be reviewed. However, instead of representing the worldsheet
matter variables with the \ads\ superspace coset 
${{PSU(2,2|4)}\over{SO(4,1)\times SO(5)}}$, the worldsheet matter
variables will be represented by the $AdS_5$ superspace coset
${{PSU(2,2|4)}\over{SO(4,1)\times SO(6)}}$ together with 
${{SO(6)}\over{SO(5)}}$ variables for $S^5$.
Although the two superspace cosets are related by a field redefinition,
the ${{PSU(2,2|4)}\over{SO(4,1)\times SO(6)}} \times 
{{SO(6)}\over{SO(5)}}$
is more convenient 
for comparing with ${\cal N}=4$ d=4
harmonic superspace since the 
${{SO(6)}\over{SO(5)}}$
variables transform 
under ${\cal N}=4$ d=4 supersymmetry
in the same manner
as harmonic variables.

\subsec{Worldsheet variables}

The $AdS_5$ superspace contains 5 bosonic variables denoted
$[x^m, z]$ for $m=0$ to 3, and 32 fermionic variables denoted
$[\t^{\a j},\tb^\ad_j,\psi^\a_j, \bar\psi^{\ad j}]$ for
$(\a,\ad)=1$ to 2 and $j=1$ to 4.
These variables appear in the 
${{PSU(2,2|4)}\over{SO(4,1)\times SO(6)}}$ supercoset as
\eqn\superc{g = \exp(x^m P_m + i\, \t^{\a j} q_{\a j} + i \, \tb_{\ad j} \bar q^{\ad j})
~\exp(i \, \psi^\a_j s_\a^j + i \, \bar\psi_{\ad}^j \bar s^{\ad}_j)~z^D \, ,} 
where $[P_m,q_{\a j},\bar q_\ad^j]$ are the ${\cal N}=4$
d=4 supersymmetry and translation generators, $D$ is the dilatation
generator, and $[s^{\a j},\bar s^\ad_j]$ are the $\cN$ d=4
superconformal generators. Under global $PSU(2,2|4)$ transformations
generated by $\Sigma$, $g$ transforms by left multiplication
as $\d g=\Sigma g$. And under
local $SO(4,1)\times SO(6)$ transformations generated by $\Omega$,
$g$ transforms by right multiplication as $\d g = g \Omega$.
Note that with the parameterization of \superc,
when $z\to 0$ at the $AdS_5$ boundary, the variables $[x,\t,\tb]$ 
transform in the usual $\cN$ d=4 superconformal manner under
global $PSU(2,2|4)$ transformations. 

 The $S^5$ space will be parameterized using a unit vector
$y^J$ for $J=1$ to 6 satisfying $y^J y^J=1$. 
Using $SO(6)$ Pauli matrices $\s^J_{jk}$, one can define
$y_{jk} = - y_{kj} = y_J \s^J_{jk}$ which satisfies the
normalization condition that ${1\over 8}\e^{jklm}y_{jk} y_{lm}=-1$.
Note that $y^{jk}  = \half \e^{jklm}y_{lm}$ and that
$y_{jk} y^{kl} =\d_j^l$.

Finally, one needs to include the left and right-moving pure spinor
variables $(\l^{\a j},\lb^\ad_j)$ and $(\lh^{\a j},\lhb^\ad_j)$,
as well as their conjugate momenta
$(w_{\a j},\wb_\ad^j)$ and $(\wh_{\a j},\whb_\ad^j)$.
These variables satisfy the pure spinor constraints
\eqn\pures
{\l^{\a j}\lb^\ad_j=0, \quad \l^{\a j}\l_\a^k - 
\half\e^{jklm}\lb_{\ad l}\lb^\ad_m=0,}
$$\lh^{\a j}\lhb^\ad_j=0, \quad \lh^{\a j}\lh_\a^k - 
\half\e^{jklm}\lhb_{\ad l}\lhb^\ad_m=0,$$
which are the four-dimensional reduction of the d=10 pure
spinor constraints
\eqn\tendpure{\l\g^M\l=0,\quad \lh\g^M\lh=0,}
for $M=0$ to 9. As in ten dimensions, gauge invariance under
$\d w = (\g^M\l) \Lambda_M$ 
and
$\d \wh = (\g^M\lh) \widehat\Lambda_M$ 
implies that
$w$ and $\wh$ can only appear in the combinations of either $SO(9,1)$
Lorentz currents $N^{MN}= {1 \over 4} (w\g^{MN}\l)$ and
$\widehat N^{MN}= {1 \over 4} (\wh\g^{MN}\lh)$, or
ghost currents $J_g= (w\l)$ and $\widehat J_g = (\wh\lh)$.
So as in ten dimensions, there are
11 independent $\l$'s and $\lh$'s and 11 gauge-invariant $w$'s and $\wh$'s.

Under the local $SO(4,1)\times SO(6)$ gauge transformations which
transform $g$ by right multiplication as $\d g=g\Omega$, one also
must transform the $y_{jk}$ and pure spinor variables.
Under the $SO(3,1)\times SO(6)$ subgroup of $SO(4,1)\times SO(6)$,
these variables transform in the obvious way as
\eqn\transfo{\d y_{jk} = c_j^l y_{lk} + c_k^l y_{jl},}
$$\d\l_{\a}^j = c_\a^\b \l_{\b}^j - c^j_k \l_{\a}^k,\quad
\d\lb^\ad_j = c^\ad_\bd \lb^\bd_j + c_j^k \lb^\ad_k,$$
$$\d w^{\a}_j = -c^{\a}_{\b} w^{\b}_j + c_j^k w^{\a}_k,\quad
\d\wb_\ad^j = -c_\ad^\bd \wb_\bd^j - c^j_k \wb_\ad^k,$$
$$\d\lh_{\a}^j = c_\a^\b \lh_{\b}^j - c^j_k \lh_{\a}^k,\quad
\d\lhb^\ad_j = c^\ad_\bd \lhb^\bd_j + c_j^k \lhb^\ad_k,$$
$$\d \wh^{\a}_j = -c^{\a}_{\b} \wh^{\b}_j + c_j^k \wh^{\a}_k,\quad
\d\whb_\ad^j = -c_\ad^\bd \whb_\bd^j - c^j_k \whb_\ad^k,$$
where $\Omega = c^j_k R^k_j -{1 \over 4}(c^\a_\b (\s^{mn})_\a^{\, \, \, \, \b}  + c^\ad_\bd 
(\bar\s^{mn})_{\, \, \, \, \ad}^\bd) M_{mn}$,
$R_j^k$ are the $SU(4)$ R-symmetry generators and $M_{mn}$ are
the $SO(3,1)$ Lorentz generators.
And under the local $SO(4,1)$ transformations which are not contained
in $SO(3,1)$, these variables transform as
\eqn\newt{\d y_{jk} =0,}
$$ \d\l_{\a}^j = - c_{\ad\a} y^{jk} \lb^{\ad}_k,
\quad \d\lb^\ad_j = c^{\a\ad} y_{jk} \l_\a^k,
 \quad \d\lh_{\a}^j = - c_{\ad\a} y^{jk} \lhb^{\ad}_k,
\quad \d\lhb^\ad_j = c^{\a\ad} y_{jk} \lh_\a^k,$$
$$
\d w^{\a}_j =  c^{\a\ad} y_{jk} \wb_{\ad}^k,
\quad \d\wb_{\ad}^j = - c_{\ad\a} y^{jk} w^{\a}_k,\quad
\d \wh^{\a}_j = c^{\a\ad} y_{jk} \whb_{\ad}^k,
\quad \d\whb_{\ad}^j =- c_{\ad\a} y^{jk} \wh^{\a}_k,$$
where $\Omega = c^{\a\ad} i \s^m_{\a\ad} M_{5m}$
and $M_{5m}$ are the four $SO(4,1)/SO(3,1)$ generators.
Note that 
\eqn\fivel{\l^{A j} = [\l_\a^j, y^{jk} \lb^{\ad}_k],\quad
w_{A j} = [w^{\a}_j, - y_{jk} \wb_{\ad}^k],}
$$
\lh^{A j} = [\lh_\a^j, y^{jk} \lhb^{\ad}_k],\quad
\wh_{A j} = [\wh^{\a}_j, - y_{jk} \whb_{\ad}^k], $$
transform covariantly as $SO(4,1)\times SO(6)$ spinors where
$A = (\a,\ad)$ is an $SO(4,1)$ spinor index. 

\subsec{Worldsheet action}

To construct the BRST-invariant worldsheet action using the
${{PSU(2,2|4)}\over{SO(4,1)\times SO(6)}}$ supercoset $g$ of \superc,
the first step is to define the left-invariant currents $J=g^{-1} \p g$
and $\bar J= g^{-1} \bar{\p} g$ taking values in the $PSU(2,2|4)$ Lie
algebra. The bosonic currents will be denoted as
$(J^m,J^5)$ for the $AdS_5$ translations, $[J^{\a\b},J^{\ad\bd},J^{\a\bd}]$
for the $SO(4,1)$ rotations, and $J^j_k$ for the $SO(6)$ rotations.
And the fermionic currents will be denoted as $[J^{j\a}, J_{\ad j}]$
for the supersymmetries and $[J_j^\a, J^j_\ad]$ for the
superconformal transformations. So the left-invariant current is
\eqn\expc{g^{-1}\p g = J^m {1 \over 2}(P_m + K_m) + J^5 D+ J^{AB} M_{AB} + J^k_j R^j_k 
+ J^{\a j} q_{\a j} + J_{\ad j} \bar q^{\ad j} + J^\a_j s_\a^j + J_\ad^j \bar s^\ad_j}
where $M_{AB} = M_{BA}$ are the $SO(4,1)$ generators, $A=(\a,\ad)$ is
an $SO(4,1)$ spinor index, and $K_m$ is the generator of special conformal transformations.

In terms of these currents, the ghost-independent contribution to the
worldsheet action is 
\eqn\matter{S_{matter} = \int d^2 z [\half \eta_{mn} J^m \bar J^n + \half J^5 \bar J^5
- {1 \over 8} (\nabla y)_{jk} (\bar\nabla y)^{jk}} 
$$
-2 J^{\a}_j \bar J_\a^j -2  J^{\a j} \bar J_{\a j} -2 J_{\ad}^j \bar J^{\ad}_j -2 J_{\ad j} \bar J^{\ad j}$$
$$ -
y_{jk} J^{\a j} \bar J_\a^k -y^{jk} J^\a_j \bar J_{\a k}
+ y^{jk} 
J_{\ad j} \bar J^\ad_k +y_{jk} J_\ad^j \bar J^{\ad k}],$$
where $(\nabla y)_{jk} = \p y_{jk} - J^l_j y_{lk} - J^l_k y_{jl}$.
The easiest way to verify this action is to use the $SO(6)$
gauge invariance to gauge-fix $y_{jk} = (\s^6)_{jk}$ where
$(\s^J)_{jk}$ for $J=1$ to 6 are the $SO(6)$ Pauli matrices.
One can then compare \matter\ with
the action written in terms of
the ${{PSU(2,2|4)}\over{SO(4,1)\times SO(5)}}$ supercoset.

When $y_{jk}=(\s^6)_{jk}$, the term $-{1 \over 8}(\nabla y)_{jk} (\bar\nabla y)^{jk}$
reduces to ${1 \over 2} \sum_{J=1}^5 J^{6J} \bar J^{6J}$ where 
$J^{6J}\equiv {1 \over 2} (\s^{6J})^{\, \, \, \,j}_k J^k_j$.
Furthermore, the third line of \matter\ reduces to 
\eqn\gfive{-(\s^6)^{jk} \ve_{AB} J^A_j \bar J^B_k -(\s^6)_{jk} \ve_{AB}
J^{A j} \bar J^{B k} }
where $\ve_{AB}$ is the $SO(4,1)$-invariant antisymmetric metric,
i.e. $\ve_{\a\bd}=0$, $\ve_{\a\b}=\e_{\a\b}$ and $\ve_{\ad\bd}=\e_{\ad\bd}$.
In this gauge,
one can easily show the equivalence of \matter\ with the ghost-independent
contribution to the action written in terms of the 
${{PSU(2,2|4)}\over{SO(4,1)\times SO(5)}}$ supercoset which is
\eqn\classtwo{S_{matter}=
\int d^2 z[ \half \eta_{ab} J_{(2)}^a \bar J_{(2)}^b - \half  
\ve_{AB}(\s^6)_{JK}
(J_{(1)}^{A J} \bar J_{(3)}^{B K}
+\bar J_{(1)}^{A J} J_{(3)}^{B K})}
$$ + {1\over 4} 
\ve_{AB}(\s^6)_{JK}
(J_{(1)}^{A J} \bar J_{(3)}^{B K} - \bar J_{(1)}^{A J} J_{(3)}^{B K})]$$
where we have used the notation 
\eqn\not{\ve_{AB} (\s^6)_{JK} J_{(1)}^{A J} \bar J_{(3)}^{B K} =  (\s^6)_{i j}
\e_{\a \b} J_{(1)}^{\a i} \bar J_{(3)}^{\b j} - (\s^6)^{i j}
\e_{\ad \bd} J_{(1) i}^{\ad} \bar J_{(3) j}^{\bd}}
and similar for the other terms. In \classtwo\
 $a=0$ to 9 is
an $SO(4,1)\times SO(5)$ vector index and the currents
$[J_{(1)}^{\a j}, J_{(1) j}^{\ad} , J_{(2)}^a, J_{(3)}^{\a j}, J_{(3) j}^{\ad}]$ are related to the currents
of \expc\ by 
\eqn\currel{J_{(1)}^{\a j} = \sqrt{2} \, J^{\a j} + \sqrt{2} \, (\s^6)^{j i} J_i^{\a},\quad
J_{(3)}^{\a j} = - \sqrt{2} J^{\a j} + \sqrt{2} (\s^6)^{j i} J_i^{\a},}
$$ J_{(1) j}^{\ad} = - \sqrt{2} \, J^{\ad}_j + \sqrt{2} \, (\s^6)_{j i} J^{i \ad},\quad
J_{(3) j}^{\ad} = -\sqrt{2} J^{\ad}_j - \sqrt{2} (\s^6)_{j i} J^{\ad i}, $$
and $ J_{(2)}^a = [J^m, J^5, J^{6J}]$.

Finally, the ghost-dependent contribution to the action is given
by 
\eqn\ghostd{S_{ghost}=\int d^2 z [w_{Aj}(\bar\nabla\l)^{Aj} -
\wh_{Aj}(\nabla\lh)^{Aj}
+ {1 \over 2} y^{jl} (\bar\nabla y)_{lk} w_{Aj}\l^{Ak} - {1 \over 2} 
y^{jl} (\bar\nabla y)_{lk} \wh_{Aj}\lh^{Ak} }
$$-2 N_{mn} \widehat N^{mn} - 4 (y^J N_{J m})(y_K \widehat N^{K m})
+ 2 N_{JK} \widehat N^{JK} - 4 (y^L N_{L J})(y_M \widehat N^{MJ}) ],$$
where in the first line of \ghostd,
$\l^{Aj}$, $w_{Aj}$, $\lh^{Aj}$ and $\wh_{Aj}$ are the $SO(4,1)\times SO(6)$ spinors
defined in \fivel\ and
\eqn\cov{w_{Aj}(\bar\nabla\l)^{Aj} = w^{\a}_j\bar\p\l_{\a}^j + \bar w_{\ad}^k y_{k i} \bar\p(y^{i j} \l_j^{\ad})
-w^{\a}_j \bar J_{\a}^{\b} \l_{\b}^j - \bar w_{\ad}^k \bar J^{\ad}_{\bd} \bar\l^{\bd}_k}
$$ + 2 w^{\a}_j \bar J_{\a \ad} y^{jk} \bar\l_k^{\ad}-2\bar w_{\ad}^k y_{kl} \bar J^{\ad \a} \l_{\a}^l
+ w^{\a}_j \bar J^j_k \l^k_{\a} + \bar w^{k}_{\ad} y_{k i} \bar J^i_m y^{m p} \bar\l_p^{\ad} ,$$
and similar for $\wh_{Aj}(\nabla\lh)^{Aj}$.
In the second line of \ghostd,
the $SO(9,1)$ Lorentz currents $N^{MN}= {1 \over 4} (w\g^{MN}\l)$ for $M=0$ to 9
are constructed
out of the $SO(3,1)\times SO(6)$ spinors $(\l^{\a j},\lb^\ad_j)$
and $(w_{\a j},\wb_\ad^j)$ and 
have been decomposed into their $SO(3,1)\times SO(6)$ components as
$[N^{mn}, N^{mJ}, N^{JK}]$.

This ghost contribution can be verified by choosing
the gauge $y_{jk}=(\s^6)_{jk}$
and comparing with
the ghost contribution 
using the ${{PSU(2,2|4)}\over{SO(4,1)\times SO(5)}}$ supercoset
which is
\eqn\ghostold{S_{ghost}=\int d^2 z
 [ w_{Aj}\widetilde{\bar\nabla}\l^{Aj} - \widehat w_{Aj}\widetilde
\nabla\lh^{Aj}
+ R_{abcd} \widetilde N^{ab} \widehat {\widetilde N}^{cd} ], }
where
$\widetilde\nabla$ only involves the $SO(4,1)\times SO(5)$ connection,
$\widetilde N^{ab}$ is constructed out of the $SO(4,1)\times SO(5)$ spinors
$\l^{Aj}$ and $w_{Aj}$,
and $R_{abcd}$ is the \ads\ curvature, i.e. $R_{abcd}=\eta_{a[d}\eta_{c]b}$
if $(a,b,c,d)$ are on $AdS_5$ and
$R_{abcd}=-\eta_{a[d}\eta_{c]b}$ if $(a,b,c,d)$ are on $S^5$.
Since 
$(\bar\nabla\l)^{Aj} = 
\widetilde{\bar\nabla}\l^{Aj}  - {1 \over 2} \bar J^{6J} (\s_{6J})^{\, \, \, \,j}_k \l^{Ak}$, the first line
of \ghostd\ reproduces the first two terms of \ghostold\ when 
$y_{jk}=\s^6_{jk}$. And by writing the $SO(4,1)\times SO(5)$ Lorentz
spinors in terms of $SO(3,1)\times SO(6)$ spinors, one finds that
when $y_{jk}=\s^6_{jk}$, the second line of \ghostd\ reduces to
the last term of \ghostold.
 
\subsec{BRST operator}

Physical closed string states in the pure spinor formalism are described
by the cohomology at $+2$ ghost number of the sum of the left
and right-moving BRST operator. In terms of the 
${{PSU(2,2|4)}\over{SO(4,1)\times SO(5)}}$ supercoset, the 
BRST operator
in an \ads\ background is 
\eqn\brstold{Q=\int d{z} ~\l^{AJ} \ve_{AB}(\s^6)_{JK} J_{(3)}^{BK} - 
\int d{\bar z} ~\lh^{AJ} \ve_{AB}(\s^6)_{JK} \bar J_{(1)}^{BK}.} 
Using the relation of \currel\ for the fermionic currents, one therefore finds
that the BRST operator in terms of the
${{PSU(2,2|4)}\over{SO(4,1)\times SO(6)}}$ supercoset is
\eqn\brstnew{Q = \int d{z}~ [\l^{\a j} (\sqrt{2} J_{\a j} - \sqrt{2} y_{jk} J_\a^k)
-\lb_{\ad j} (\sqrt{2} J^{\ad j} + \sqrt{2} y^{jk} J^\ad_k)]}
$$- \int d{\bar z}~ [\lh^{\a j} (\sqrt{2} \bar J_{\a j} +\sqrt{2} y_{j k} \bar J^k_{\a})
+\lhb_{\ad j} (\sqrt{2} \bar J^{\ad j} - \sqrt{2} y^{jk} \bar J^\ad_k)].$$

Under a BRST transformation of \brstold\ a representative $g'$ of the supercoset 
${{PSU(2,2|4)}\over{SO(4,1)\times SO(5)}}$ transforms as       


\eqn\mudanca{ \d g' = g' (\l^{\bar{\a}} T^1_{\bar{\a}} + \hat{\l}^{\hat{\bar{\a}}} T^3_{\hat{\bar{\a}}}) }

\noindent where $T^1_{\bar{\a}}$ and $T^3_{\hat{\bar{\a}}}$ are the fermionic generators 
of the $PSU(2,2|4)$ , $\bar{\a} = 1, \ldots, 16$ and 
$\hat{\bar{\a}}=1,\ldots,16$. These generators are related with $q_{\a i}, \, q^i_{\ad} , \,
s^j_{\a} , \, s^{\ad}_j $ as   

\eqn\algebra{ T^1_{\a i} = {\sqrt{2} \over 4} q_{\a i} - {\sqrt{2} \over 4} (\s^{6})_{i j} s^{j}_{\a} ,
\quad  T^{1 i}_{\ad} = - {\sqrt{2} \over 4} q^i_{\ad} - {\sqrt{2} \over 4} (\s^{6})^{i j} s_{\ad j} }
$$ T^{3}_{\a i} = - {\sqrt{2} \over 4} q_{\a i} -{\sqrt{2} \over 4} (\s^{6})_{i j} s^j_{\a} , \quad
T^{3 i}_{\ad} = - {\sqrt{2} \over 4} q^{i}_{\ad} + {\sqrt{2} \over 4} (\s^6)^{i j} s_{\ad j} $$

Note that in the gauge $y_{jk} = (\s^6)_{jk}$, \currel\ and \algebra\ satisfy
\eqn\coeffs{ J^{\a j} q_{\a j} + J_{\ad j} q^{\ad j} + J^{\a}_j s^j_{\a} + J^{j}_{\ad} s^{\ad}_j
 = J^{1 \bar{\a}} T_{\bar{\a}} + J^{3 \hat{\bar{\a}}} T_{\hat{\bar{\a}}} }

Using the relations above we can see  
that the BRST transformation of $g$  under \brstnew\  is 
\eqn\brstg{\d g = g {\sqrt{2} \over 4}[i \l^{+ \a j}  q_{\a j} -
i \l^{- \a j} y_{jk} s_\a^k + i
\lb^{+}_{\ad j} \bar q^{\ad j} + i 
\lb^-_{\ad j} y^{jk} \bar s^{\ad}_k ]}
where $(q^\a_j,\bar q_\ad^j)$ and $(s_\a^j,\bar s^\ad_j)$ are the
$\cN$ d=4 supersymmetries and superconformal transformations and 
\eqn\define{\l^{- \a j} \equiv - i (\l^{\a j} + \lh^{\a j}), \quad \l^{+ \a j} \equiv - i (\l^{\a j} - \lh^{\a j})}
$$ \lb^{-}_{\ad j} \equiv i (\lb_{ \ad j} - \lhb_{\ad j}), \quad 
\lb^{+}_{\ad j} \equiv i (\lb_{ \ad j} + \lhb_{\ad j}). $$

Since $w_{Aj}$ and $\wh_{Aj}$ are
conjugate to $\l^{Aj}$ and $\lh^{Aj}$, one finds
that $w_{Aj}$ and $\wh_{Aj}$ transform under BRST as
\eqn\wtr{\d w_{\a j} = \sqrt{2} J_{\a j} - \sqrt{2} y_{jk} J_\a^k,\quad
\d \wb^{\ad j} =- \sqrt{2} J^{\ad j} - \sqrt{2} y^{jk} J^\ad_k,}
$$\d \wh_{\a j} = \sqrt{2} \bar J_{\a j} + \sqrt{2} y_{jk} \bar J_\a^k,\quad
\d \whb^{\ad j} = \sqrt{2} \bar J^{\ad j} - \sqrt{2} y^{jk} \bar J^\ad_k.$$
Note that these BRST transformations are defined up to the gauge
transformation $\d w = (\g^M \l) \Lambda_M$ and
$\d \wh = (\g^M \lh) \widehat\Lambda_M$. 


Finally, the BRST transformations of $\l^{Aj}$, $\lh^{Aj}$ and $y^{jk}$
are zero. However, note that these variables transform under local
$SO(4,1)\times SO(6)$ transformations. So if the BRST transformation
on $g$ needs to be compensated by a local $SO(4,1)\times SO(6)$
transformation in order to preserve a gauge-fixing condition, these variables
will transform under the compensating $SO(4,1)\times SO(6)$ gauge
transformation.

\newsec{Cohomology Analysis}

In this section, the BRST operator in an \ads\ background
will be expanded in terms of the $AdS_5$ radial variable $z$.
The cohomology will then be shown to be described by the boundary
value of the state near $z=0$.

To simplify the analysis of cohomology, it will be convenient
to express the BRST operator
in terms of the worldsheet variables $[x,\t,\psi,z,y,\l, \lh]$
and their canonical momenta $[P_x,P_\t, P_\psi, P_z, P_y, P_\l, P_\lh]$
instead of the worldsheet variables and their time derivatives. 
As usual, the canonical momenta will be defined as $P_x= {{\p L}\over
{\p (\p_\tau x)}}$, etc. Note that unlike in the Green-Schwarz formalism
which has first and second-class constraints, there are no constraints
on the canonical momenta in the pure spinor formalism. 
Using the Lagrangian of \matter\ and \ghostd, one finds, for example, that 
\eqn\momemc{P_z = -{1 \over 4z} ({\dot z \over z} + 2 \dot\t^{\a j} \psi_{\a j} + 2 \dot{\bar\t}_{\ad j} 
\bar \psi^{\ad j})}
where dot means derivative with respect to $\tau$.




Expanding the sum of the left
and right-moving BRST operator given in \brstnew\ in terms of these variables and $z$,
one finds that the BRST operator splits as
$Q = Q_{-\half} + Q_{\half} + Q_{3\over 2} + ... $ where
\eqn\splitq{Q_{-\half} = ( {\sqrt{2} \over 2 }) z^{-\half}
\l^{+ \g m} y_{m i} P_{\psi^\g_i} - ({\sqrt{2} \over 2 }) z^{-\half}  
\lb^{+ \ad}_j
y^{j i} P_{\bar\psi^{i \ad}} + f(\del_{\s} X)}
$$Q_{\half} = ( {\sqrt{2} \over 2 }) z^\half 
\l^{- \a i}(- P_{\t^{\a i}} 
- i (\s^a)_{\a \ad} \bar \t^{\ad}_i 
P_{x^a} + 2 \psi_{\a i} z P_z +4 \psi_{\a k} P_{y^{i j}} y^{j k} -  \psi_{\a i} P_{y^{jk}}y^{k j}  
 $$
$$ -4 \psi_{\a k} \l^k_\b P_{\l^i_\b} + \psi_{\a i} \l_{\b}^k P_{\l_\b^k} 
- 4 \psi_{\a k} \lh^k_{\b} P_{\lh^i_{\b}} + \psi_{\a i} \lh^k_{\b} P_{\lh^k_\b}
- 2 \psi^{\b}_i \l^j_\b P_{\l^{\a j}} - 2 \psi_{\b i} \l^j_\a P_{\l^j_\b} $$
$$ - 2 \psi^{\b}_i \lh^j_\b P_{\lh^{\a j}} - 2 \psi_{\b i} \lh^j_\a P_{\lh^j_\b}
+4 \psi_{\a k} \bar\l^{\ad}_i P_{\bar\l^{\ad}_k} - \psi_{\a i} \bar\l^{\ad}_k P_{\bar\l^{\ad}_k}
+4 \psi_{\a k} \lhb^{\ad}_i P_{\lhb^{\ad}_k} - \psi_{\a i} \lhb^{\ad}_k P_{\lhb^{\ad}_k}$$
$$ + 4 \psi_{\a j} \psi^{\b}_i P_{\psi^{\b}_j} - 2 \psi_{\a j} \bar \psi^j_\ad P_{\bar\psi^i_\ad}) + ({\sqrt{2}
\over 2 }) z^\half 
\lb^{- \ad}_i (- P_{\bar\t^\ad_i} - i \t^{\b i} (\s^a)_{\b \ad} P_{x^a} -2 \bar\psi^i_\ad 
z P_z $$
$$ -4 \bar\psi^l_{\ad} P_{y_{ij}}y_{jl} +  \bar\psi^i_{\ad} P_{y_{k j}} y_{j k} 
-4 \bar\psi^j_{\ad} \l^{i}_\a P_{\l^j_\a} + \bar\psi^i_\ad \l^k_\a P_{\l^k_\a} 
-4 \bar\psi^j_{\ad} \lh^{i}_\a P_{\lh^j_\a} + \bar\psi^i_\ad \lh^k_\a P_{\lh^k_\a} $$
$$ + 4 \bar\psi_{\ad}^l \bar\l^\bd_l P_{\bar\l^\bd_i} - \bar\psi^i_\ad \bar\l^\bd_k P_{\lb^\bd_k} 
+ 4 \bar\psi_{\ad}^l \lhb^\bd_l P_{\lhb^\bd_i} - \bar\psi^i_\ad \lhb^\bd_k P_{\lhb^\bd_k} 
-2 \bar\psi^i_{\bd} \lb^{\bd}_k P_{\lb^{\ad}_k}+ 2 \bar\psi^{\bd i}\lb_{\ad k} P_{\lb^\bd_k}$$
$$ -2 \bar\psi^i_{\bd} \lhb^{\bd}_k P_{\lhb^{\ad}_k}+ 2 \bar\psi^{\bd i}\lhb_{\ad k} P_{\lhb^\bd_k}
-4 \bar\psi^m_\ad \bar\psi^{i \bd} P_{\bar\psi^{m \bd}} - 2 \psi_{\a k} \bar\psi^k_\ad P_{\psi_{\a i}}) $$
$$ + ({\sqrt{2} \over 2}) 2 z^{\half} 
\l^{+ \g m} y_{m i} ( \bar\psi^i_\ad \lb_k^\ad y^{k j} P_{\l^{\g j}} 
+ \bar\psi^{i \ad} y_{k l} \l^l_\g P_{\lb^\ad_k} +  \bar\psi^i_\ad \lhb^{\ad}_k y^{k j} P_{\lh^{j \g}}
+  \bar\psi^{i \ad} y_{k l} \lh^l_\g P_{\lhb^{\ad}_k}) $$
$$ - ({\sqrt{2} \over 2}) 2 z^{\half} 
\lb^{+ \ad}_m y^{ m i} (  \psi_{\a i} \lb_{k \ad}
y^{j k} P_{\l^j_\a} +  \psi^\b_i \l_\b^l y_{k l} P_{\lb^{\ad}_k}
+ \psi_{\a i} \lhb_{\ad k} y^{j k} P_{\lh^j_\a} +  \psi^\a_i \lh^l_\a y_{k l} P_{\lhb^{\ad}_k} )  
+ g(\del_{\s} X) $$ 
$$ 
Q_{3\over 2}= i ({\sqrt{2} \over 2}) 
z^{{3 \over 2}} 
\l^{+ \g m} y_{m i}  (\s^a)_{\g \ad} 
\bar\psi^{i \ad} P_{x^a} -  i ({\sqrt{2} \over 2}) 
z^{{3 \over 2}} 
\lb^{+ \ad}_m y^{ m i}
 \psi^\b_i (\s^a)_{\b \ad} P_{x^a}  
+ ...$$
and $...$ are terms which are at least quadratic in $\psi$. We have suppressed
the $\int d \s$ in the expression above and $f(\del_{\s} X)$ and $g(\del_{\s} X)$ denote additional
terms that contain sigma derivatives of the fields.   
We will work to lowest order in $\alpha'$ so possible normal-ordering
contributions to $Q$ will be ignored.






Near the $AdS_5$ boundary, physical states $V$ can be expanded as
$V = \sum_{d\geq d_0} V_d$ where $V_d$ is proportional to $z^d$
and $V_{d_0}$ is the leading behavior
near $z=0$. Defining the degree to be the power of $z$ in the expansion,
$V$ has a 
%
minimum degree $d_0$, so one can use
standard methods to compute the cohomology of $Q$. 
One first
computes the cohomology of 
$Q_{-\half}$, then computes the cohomology of $Q_\half$ restricted
to states in the cohomology of $Q_{-\half}$, 
then computes the cohomology of $Q_{3\over 2}$ restricted
to states in the cohomology of $Q_{-\half} + Q_\half$, etc.
This procedure is
well defined since the complete BRST operator given in \brstnew\ is nilpotent,
and after performing the $z$ expansion, this implies  
$\{ Q_{\half} , Q_{\half} \} + 2 \{ Q_{-\half} , Q_{3 \over 2} \} = 0.$
So $Q_{\half}$ is a nilpotent operator when acting on states 
in the cohomology of $Q_{-\half}$. The same argument of nilpotency applies to $Q_{3 \over 2} \, , \,   
Q_{5 \over 2} \, , \, \ldots$. 

In what follows we will focus on the zero mode BRST cohomology which is
relevant for the supergravity states.  
Because of the usual quartet argument, the zero mode cohomology of 
\eqn\cohomqhalf{Q_{-\half}=
 ( {\sqrt{2} \over 2 }) z^{-\half} 
\l^{+ \g m} y_{m i} P_{\psi^\g_i} - ({\sqrt{2} \over 2 }) z^{-\half} 
\lb^{+ \ad}_j
y^{j i} P_{\bar\psi^{i \ad}} }
will be assumed \foot{
As shown by A. Mikhailov and R. Xu in \mikh, this treatment using the quartet 
argument is too naive and there is one state at ghost-number two
in the $Q_{-\half}$ cohomology that depends on $\l^+$ which is   
$(\l^{+} \g^M \hat{\psi})(\l^{-} \g_M \hat{\psi})$ \MikhailovAF.
However, if one
allows dependence on the non-minimal
pure spinor variables $\lti$ and $r$ described in the following section, 
this state in the cohomology of $Q_{-\half}$
can be represented by 
$(\l^- \lti)^{-2}
(\l^-\g^M\hat\psi)
(\l^-\g^N\hat\psi)
(\l^-\g^P\hat\psi) (r \g_{MNP} \lti)$
which is independent of $\l^+$.
} 
to be given by states which
are independent of $\l^+$ 
and which only
depend on $\psi$ in the combination $\l^{-}\g^M\hat\psi$ 
where $\hat\psi \equiv y_J (\g^J \psi)$.
This combination
$\l^{-}\g^M\hat\psi$ has been written in ten-dimensional notation where
$\l^{-\aba}$ and $\hat\psi^{\aba}$ are $d=10$ Weyl spinors, $\aba=1$ to 16, and
$M=0$ to 9.
Note that the pure spinor condition \tendpure\ implies that 
\eqn\pureminus{\l^-\g^M\l^+=0 \quad{\rm and} \quad 
\l^+\g^M\l^+ + \l^-\g^M\l^- =0,}
so that $Q_{-\half}(\l^{-}\g^M\hat\psi)=0$.


Since states in the cohomology of $Q_{-\half}$ are independent of
$\l^+$, the condition 
$\l^+\g^M\l^+ + \l^-\g^M\l^-=0$ implies that $\l^{-}\g^M\l^{-}=0$,
i.e. $\l^{-}$ is a pure spinor with 11 independent components.
This implies that $\l^{-}\g^M\hat\psi$ has only 5 independent components.
Therefore, states in the cohomology of $Q_{-\half}$ depend on
the 21 bosonic coordinates
$[x,z,y,\l^-]$ and 21 fermionic coordinates $[\t,\l^-\g^M\hat\psi]$.

The next step is to compute the cohomology of $Q_{\half} + Q_{3\over 2} + ...$
when restricted to states in the cohomology of $Q_{-\half}$. Since states
in the cohomology of $Q_{-\half}$ are independent of $\l^+$ and only depend
on $\psi$ in the combination $\l^- \g^M \hat\psi$,
any terms proportional to $\l^{+}$ in
$Q_{3\over 2} + 
...$ act as zero when restricted to these states. 
And any terms proportional to $\l^{-}$ in 
$Q_{3\over 2} + ...$ 
are at least cubic in $\psi$ and must act as zero when restricted to these 
states. This is because any operator which is linear in $\l^-$ and cubic
in $\psi$ cannot be expressed in terms of the five $\l^-\g^M\hat\psi$ variables.
Thus all the operators in $Q_{3\over 2} + ...$ act
as zero when restricted to these states.
So
computing the cohomology of $Q$ reduces to computing the cohomology
of $Q_\half$ restricted to states depending on the variables
$[x,\t,z,y,\l^-,\l^-\g^M\hat\psi]$. 



Since $Q_\half$ has a fixed degree, one only needs to consider
vertex operators of a fixed degree to compute its cohomology.
If $V$ is the original vertex operator in the cohomology of $Q$,
this will be the term $V_{d_0}$ of lowest degree in $V$ after
restricting to states in the cohomology of $Q_{-\half}$.
For this reason, the vertex operator inside the region of
validity of the $z$ expansion is determined up to a BRST-trivial
quantity by its boundary value $V_{d_0}$. 
Holography predicts that $V_{d_0}$ should be dual to a gauge-invariant
super-Yang-Mills operator, and the precise relation will
be discussed in the next 
section for the case of supergravity vertex operators which are dual to
half-BPS super-Yang-Mills operators. 

\newsec{Half-BPS States}

\subsec{BRST cohomology}

In this section, the zero mode BRST cohomology at $+2$ ghost
number will be related to the
dual of half-BPS gauge-invariant $d=4$ ${\cal N}=4$ super-Yang-Mills operators. 
Zero mode cohomology at $+2$ ghost number corresponds to supergravity
states which, like the dual half-BPS super-Yang-Mills operators, will
be expressed using 
${\cal N}=4$ d=4 harmonic superspace.

As argued in the previous section, the BRST cohomology is described
by states in the cohomology of $Q_{\half}$ depending on 
$[x,z,y,\t, \l^{-},\l^{-}\g^M\hat{\psi}]$ where $\l^{-}\g^M\l^{-}=0$. 
In what follows we are going to suppress the minus superscript in $\l^-$.
The zero mode contribution
to $Q_\half$ written in ten dimensional notation is:
\eqn\zeroqhalf{Q_{\half} = z^{1 \over 2} [\l^{\aba} D_{\aba} + 4 (\l\g^{jk}\hat{\psi})
{\p\over{\p y^{j k}}} +  y_{i j} (\l\g^{i j} \hat{\psi})(2 z {\p\over{\p z}} +
 y^{m t}{\p\over{\p y^{m t}}} - \l^{\aba} {\p\over{\p \l^{\aba}}})] + \widetilde{w}^{\aba} r_{\aba}}
where $D_{\aba} = -{\p\over{\p\t^{\aba}}} - (\t \g^m)_{\aba} {\p\over{\p x^m}}$
is the d=4 dimensional reduction of the d=10 supersymmetric derivative.

In  \zeroqhalf\ we have included the usual non-minimal pure spinor term 
 $\widetilde{w}^{\aba} r_{\aba}$ \BerkovitsBT\
that is not present in \splitq\  . The inclusion of this additional term is necessary 
\foot{Although some expressions for vertex operators in the next subsection 
will depend on $\lti_{\aba}$ and 
$r_{\aba}$, it should be noted that there always exists a gauge in which the 
vertex operator depends only on minimal variables. This is clear from
the expression of \purever. However, to express the vertex operator in
terms of harmonic superspace variables, dependence on non-minimal
variables appears to be necessary when the supergravity state is dual to
a half-BPS state involving four or more super-Yang-Mills fields. }   
because, as
will be seen below,  
some of the results can be expressed
as a function of $(\l \g^M \hat{\psi})$ only after introducing the
non-minimal bosonic pure spinor variables $\lti_{\aba}$ satisfying     
\eqn\tildeprop{\lti_{\aba} (\g^M)^{\aba \bba} \lti_{\bba} = 0 \; .} 
The $\widetilde{w}^{\aba}$ are the conjugate momenta of $\lti_{\aba}$
which act on functions of $\lti_{\aba}$
as $\p\over{\p{\lti_{\aba}}}$, and $r_{\aba}$ is a fermionic spinor which satisfies
\eqn\rprop{\lti_{\aba} (\g^M)^{\aba \bba} r_\bba = 0.}  

To rewrite $Q_\half$ given in \splitq\ in the concise form of \zeroqhalf,
we have performed a few manipulations.
Firstly,
we have redefined $\l$ in order to adsorb the overall factor of ${ \sqrt{2} \over 2}$.
Secondly, the term  
$z^{\half} 4 (\l \g^{jk} \hat{\psi}) {\p \over {\p y^{j k}}}$ in \zeroqhalf\ is 
understood to not act on $(\l \g^M \hat{\psi})$ even though $(\l\g^M\hat{\psi})$
depends on $y_{i j}$.
This is the case because we have not explicitly included in \zeroqhalf\
the terms 
$$z^{1 \over 2}[- 2 \l^{\a i} \psi_{\a j} \bar{\psi}^{j}_{\dot{\a}} 
P_{\bar\psi^i_{\dot{\a}}}
- 2 \bar{\l}_i^{\dot{\a}} \psi_{\a k} \bar{\psi}^k_{\dot{\a}} P_{\psi_{\a i}}
 +4 \l^{ \a i} \psi_{\a j} \psi^{\b}_i P_{\psi^{\b}_j}-
4 \lb^{ \ad}_i \bar\psi^{m}_{\ad} \bar\psi^{i \bd} P_{\bar\psi^{m \bd}}]$$
from \splitq, and one can 
show using pure spinor conditions for $\l$ that
\eqn\elimination{\Big{(}4(\l \g^{jk} \hat{\psi}){\p \over {\p y^{j k}}} 
- 2 \l^{\a i} \psi_{\a j} \bar{\psi}^{j}_{\dot{\a}} P_{\bar\psi^i_{\dot{\a}}}  
- 2 \bar{\l}_i^{\dot{\a}} \psi_{\a k} \bar{\psi}^k_{\dot{\a}} P_{\psi_{\a i}}}
$$ +4 \l^{ \a i} \psi_{\a j} \psi^{\b}_i P_{\psi^{\b}_j} -
4 \lb^{ \ad}_i \bar\psi^{m}_{\ad} \bar\psi^{i \bd} P_{\bar\psi^{m \bd}} 
\Big{)}(\l \g^M \hat{\psi}) = 0 .$$
Finally, to
extract the term $ -z^{\half} y_{i j} (\l \g^{i j} \hat{\psi}) \l^{\aba} {\p \over {\p \l^{\aba}}} $ 
in \zeroqhalf\ from the complicated dependence of $Q_\half$ in 
\splitq\ in ${\p \over {\p \l}}$, we have used that terms with  
$\l^{+}$ act as zero on the states in the cohomology of  
$Q_{-\half}$ and we have omitted terms that are zero by the pure
spinor condition such as 
$$ \l^{- \a i} (4 \psi_{\a k} \lb^{\ad}_i P_{\lb^{\ad}_k} 
+ 4 \psi_{\a k} \hat\lb^{\ad}_i P_{\hat\lb^{\ad}_k}) f(\l^{- \aba}) = 0.$$
   


\subsec{Supergravity vertex operators}

It will now be shown that supergravity
vertex operators in the zero mode cohomology which
are proportional to $y_{J_1} ... y_{J_{N-1}}$ are related to half-BPS
operators that are constructed from $N$ super-Yang-Mills fields. 
We start the analysis of the cohomology of the operator $Q_{\half}$ 
considering supergravity vertex operators which are independent
of $y_{jk}$. 
These states must be annihilated by 
$2z {\p\over{\p z}} -
 \l {\p\over{\p \l}} $ and must be in the $+2$ ghost-number cohomology
of $Q= \l^\aba D_\aba$. Since $\l^\aba D_\aba$ is the four-dimensional reduction
of the ten-dimensional BRST operator $Q= \l^\aba D_\aba$, these states
are the antifields of super-Yang-Mills described by the bispinor
superfield $A^*_{\aba \bba}(x,\t)$ \BerkovitsRB . 
In other words, the $y$-independent
supergravity vertex operators are
\eqn\yind{V = z \l^\aba \l^\bba A^*_{\aba\bba}(x,\t)} 
where $A^*_{\aba\bba}(x,\t)$ is the dimensional reduction of the ten-dimensional
super-Yang-Mills antifield. 
Note that the factor of $z$ is required to cancel the BRST transformation
of $\l$ and implies that $V$ carries zero dimension (since
$A^*_{\aba\bba}$ carries dimension $+1$ in units where $x^m$ and $z$ carry dimension
$-1$).
At zero momentum, the vertex operator of \yind\ can be gauged to
\eqn\zeroanti{z\l^\aba \l^\bba A^*_{\aba\bba} =z[ (\l\g^M\t)(\l\g^N\t)(\t\g_{MN})^\aba
\psi_\aba^* + }
$$
(\l\g^M\t)(\l\g^N\t)(\t\g_{MN p}\t)a^{*p} +
(\l\g^M\t)(\l\g^N\t)(\t\g_{MN}^{\quad \, \, \, [jk]}\t)\phi^{*}_{jk}]  $$
where $a^*_p$, $\phi^*_{jk}$ and $\psi^*_\a$ are the 
antifields to the gluon $a_m$, scalars $\phi^{jk}$, and
gluino $\psi^\a$. 
So 
these
$y$-independent operators are the duals to super-Yang-Mills ``singleton''
operators, i.e. the duals to abelian super-Yang-Mills fields.

We next consider supergravity vertex operators in the cohomology which
are linear in $y$. The simplest example is
the operator $V = i \, \l^{\a j} \l_\a^k y_{jk}$ which is linear
in $y$. Note that $V$ is real since
\eqn\purey{\l^{\a i} \, \l_{\a}^j \, y_{ij} =- \lb^\ad_{i} \, \lb_{\ad j} \, y^{ij}.}  
And $V$ is annihilated by $Q_\half$ since
$\e_{jklm}\l^{\a j}\l_\a^k \l^{\b l}=0$ and $\lb^{\ad}_j \l^{\a j}=0$ imply
that $(\l\g^{ij}\hat{\psi}) {\p\over{\p y^{ij}}} (\l\l y)=0$. Since $V$ is
a $PSU(2,2|4)$ scalar, it
corresponds to the zero-momentum dilaton that is dual
to the super-Yang-Mills action.

To construct the general supergravity vertex operator in the BRST cohomology,
recall that gauge-invariant half-BPS operators involving $N$
super-Yang-Mills field strengths are elegantly described in harmonic superspace
as \howewestn\ferr\HeslopNP
\eqn\harmint{W^{(N)}(u,x,\t) = (uu)^{i_1 j_1} ... (uu)^{i_N j_N} {\rm{Tr}}[
W_{i_1 j_1}(x,\t) ... W_{i_N j_N}(x,\t)]}
where $W_{jk}(x,\t)$ is an ${\cal N}=4$ d=4 superfield satisfying
the constraints \SohniusWK
\eqn\identidades{\nabla_{\a i} W_{jk} = \nabla_{\a [i}W_{jk]},}
$$ \bar{\nabla}^{i}_{\ad} W_{j k}= - {2 \over 3} \d^{i}_{[j} \bar{\nabla}^{l}_{\ad} W_{k]l}. $$ 
In components,
\eqn\compw{W_{jk} = \phi_{jk} + \t_{\ad [j} \xi_{k]}^{\ad}
+\t^{\a l}\xi_\a^m \e_{jklm} + \t_{j \ad} \t_{k \bd} F^{\ad\bd} +
\e_{jklm}\t^{\a l}\t^{\b m} F_{\a\b} + ...}
where $\phi_{jk}$ are the scalars, $\xi^k_\a$ and $\xi^\ad_k$ are the chiral
and antichiral gluinos, and $F_{\a\b}$
and $F^{\ad\bd}$ are the self-dual and anti-self-dual field strengths. 
The expression $(uu)^{jk}$ denotes $\e^{JK} u^j_J u^k_K$ where
$G=(u^j_J, \bar u^j_{J'})$ are harmonic variables parameterizing the
coset ${SU(4)}\over{S(U(2)\times U(2))}$ with $j=1$ to 4 and $J,J'=1$
to 2. The inverse coset will be defined as $G^{-1} = (\bar u_j^J, u_j^{J'})$
where the variables $u$ and $\bar u$ satisfy the constraints
\eqn\uconst{ u^j_J \bar u_j^K = \d_J^K,\quad
\bar u^j_{J'}  u_j^{K'} = \d_{J'}^{K'},\quad u^j_J u_j^{K'} =
0,\quad \bar u^j_{J'} \bar u_j^K =0,}
$$
(u u)^{jk} = \e^{JK} u^j_J u^k_K =\half \e^{jklm} u_l^{J'} u_m^{K'} \e_{J'K'}.$$

Using the superspace constraints, 
one finds that $W^{(N)}(u,x,\t)$ satisfies
\eqn\harmdef{ u^j_J D_{\a j} W^{(N)} = u^{J'}_j \bar{D}^j_\ad W^{(N)}=0,}
i.e. $W^{(N)}$ is a G-analytic superfield. Furthermore, since 
$W^{(N)}$ is independent of $\bar u$, it satisfies
\eqn\analdef{ (u^j_J {\p\over{\p\bar u^j_{J'}}}) W^{(N)}=0,}
i.e. $W^{(N)}$ is  an H-analytic superfield. A superfield that is both G-analytic and H-analytic
will be called an analytic superfield for short.
So if $U(1)$ charge is defined as $\half (u {\p\over{\p u}} -
\bar u {\p\over{\p \bar u}})$, half-BPS states constructed from
$N$ super-Yang-Mills field strengths 
are described by analytic superfields
of $+N$ $U(1)$ charge.

To construct the duals to these analytic superfields, consider the
superspace integral \HeslopNP
\eqn\harmint{\int d^4 x \int du \int d^8 (u\t) W^{(N)}(u,x,\t)
T^{(4-N)}(u,\bar u,x,\t)}
where $\int du$ denotes an integral over the compact space
${SU(4)}\over {S(U(2)\times U(2))} $ and using the definitions 
\eqn\nada{D^{\prime 4} = D^{\a J'} D_{\a}^{K'} D^{\b}_{J'} D_{\b K'},
\quad \bar{D}^{4}=\bar{D}_{\ad}^J \bar{D}^{\ad K} \bar{D}_{\bd J} \bar{D}^{\bd}_K,} 
where $D_{\a J'} = \bar u^j_{J'} D_{\a j}$ and $\bar{D}^{J}_{\ad} = \bar u^J_{j} \bar D^j_{\ad}$ ,  
one can write
$\int d^8 (u\t) = D^{\prime 4} \bar{D}^4$.

For the integral to be supersymmetric and non-vanishing, $T^{(4-N)}$
must be a G-analytic (but not necessarily H-analytic)
superfield of $U(1)$ charge $(4-N)$. Furthermore,
the integral of \harmint\ is invariant under the gauge transformation
\eqn\gaugeT{ \d T = (u {\p\over{\p \bar u}})_J^{J'} \Lambda^J_{J'} }
for any G-analytic superfield $\Lambda^J_{J'}$. So the dual to a half-BPS
state constructed from $N$ super-Yang-Mills fields is described 
by a G-analytic superfield $T$ of $U(1)$ charge $(4-N)$ which is defined
up to the gauge transformation of \gaugeT.

In \HeslopNP, 
these superfields $T^{(4-N)}(u, \bar{u}, x, \t)$ were related to chiral 
${\cal N}=4$ d=4 superfields coming from the \ads\ Type IIB
chiral field strength at the $AdS_5$ boundary. 
However, in this paper, the superfields $T^{(4-N)}$ will instead be related
to \ads\ Type IIB gauge superfields $A_{\bar\a\bar\b}$
which appear in the BRST-invariant supergravity
vertex operators 
$V = \l^{\bar{\a}} {\lh}^{\bar{\b}} A_{\bar{\a} {\bar{\b}}}$ of \purever.  
Near the $AdS$ boundary, 
BRST-invariant
supergravity vertex operators dual to half-BPS states constructed from
$N$ super-Yang-Mills fields
will have the form
$V=  z^{2-N} \l^\aba \l^\bba A^{(N)}_{\aba\bba}(y,x,\t, \l\g^M\hat{\psi})$
up to non-minimal variables
and the precise relation
between $V$ and $T$ is 
\eqn\brstclo{V =  z^{2-N} \int du [(y uu)^{N-1} \Omega^{(0)} T + 
8 (N-1)(y uu)^{N-2} \O^{(1)}T + }
$$ 8^2 (N-1)(N-2)(y uu)^{N-3} \O^{(2)}T+
8^3 (N-1)(N-2)(N-3)(y uu)^{N-4} \O^{(3)}T+$$
$$
8^4 (N-1)(N-2)(N-3)(N-4)(y uu)^{N-5} \O^{(4)}T] ,$$

\noindent where  
\eqn\Omegaum{\Omega^{(0)} = {1 \over 16}(\l \lti)^{-2} (\l \g^M \Dt) (\l \g^N \Dt) 
(\l \g^P \Dt) (\l \g^S \Dt) (\lti \g_{M N P S T} \lti)
\www^T}
$$ + {1 \over 2} \, z^{- {1 \over 2}} 
(\l \lti)^{-2} (\l \g^M \Dt) (\l \g^N \Dt) (\l \g^P \Dt) (r \g_{P N M} \lti)  \, , $$

\eqn\omegaone{\O^{(1)} = {1 \over 4} (\l \lti)^{-2} (\l \g^M \hat{\psi}) (\l \g^N \Dt) (\l \g^S \Dt) (\l \g^P \Dt)
(\lti \g_{M N S P T} \lti) \www^T}
$$ + {3 \over 2} \, z^{- {1 \over 2}} 
(\l \lti)^{-2} (\l \g^M \hat{\psi}) (\l \g^N \Dt) (\l \g^T \Dt) (r \g_{T N M} \lti) \, , $$

\eqn\omegatwo{\O^{(2)} = {3 \over 8}  (\l \lti)^{-2} (\l \g^M \hat{\psi}) 
(\l \g^N \hat{\psi}) (\l \g^S \Dt) (\l \g^P \Dt)
(\lti \g_{M N S P T} \lti) \www^T}
$$ + {3 \over 2} \, z^{- {1 \over 2}} (\l \lti)^{-2}(\l \g^M \hat{\psi}) (\l \g^N \hat{\psi}) 
(\l \g^S \Dt)  (r \g_{S N M} \lti) \, ,$$

\eqn\omegathree{\O^{(3)} = {1 \over 4}(\l \lti)^{-2} (\l \g^M \hat{\psi}) 
(\l \g^N \hat{\psi}) (\l \g^S \hat{\psi}) (\l \g^P \Dt)
(\lti \g_{M N S P T} \lti) \www^T}
$$ + {1 \over 2} \, z^{- {1 \over 2}} (\l \lti)^{-2}(\l \g^M \hat{\psi}) (\l \g^N \hat{\psi}) 
(\l \g^S \hat{\psi}) (r \g_{S N M} \lti) \, , $$
\eqn\omegafour{\O^{(4)} ={1 \over 16} (\l \lti)^{-2} (\l \g^M \hat{\psi}) 
(\l \g^N \hat{\psi}) (\l \g^S \hat{\psi}) (\l \g^P \hat{\psi})
(\lti \g_{M N S P T} \lti) \www^T .} 
In the above formulas,
$\www_M$ and $\wwb_M$ are null vectors with nonzero components   
defined by $\www_J = -{1 \over 4} \s_J^{jk} (uu)_{jk}$  and  $\wwb_J = -{1 \over 4} \s_J^{jk}
(\bar u\bar u)_{jk}$ where $\s_J^{jk}$ are $SO(6)$ Pauli matrices,
and $\Dt \equiv \wwb_M (\g^M D)$. 

When $N<4$, the terms in \brstclo\ do not depend on $\Omega^{(3)}$ and 
$\Omega^{(4)}$ and one can choose a gauge such that $V$ is independent of
the non-minimal variables. In this gauge, $V$ is equal to
\brstclo\ but with $\Omega^{(0)}$, $\Omega^{(1)}$ and $\Omega^{(2)}$ 
replaced
with
\eqn\omina{ \Omega^{(0)}_{min} =  
- {1 \over 4} (\l \g^M \Dt)(\l \g^N \Dt)(\Dt \g_{M N P} \Dt) \www^P
,}
$$\Omega^{(1)}_{min} = 
 - (\l \g^M \hat{\psi})(\l \g^N \Dt)(\Dt \g_{M N P} \Dt) \, \www^P
$$
$$ + 24 \, (\l \g^{M} \hat{\psi}) \wwb_M \, (\l \g^{m} \Dt) \, {\partial
\over \partial x^m} \, , $$
$$\Omega^{(2)}_{min} = 
- {3 \over 2} (\l \g^M \hat\psi)(\l \g^N \hat\psi)(\Dt \g_{M N P} \Dt) \www^P
$$
$$ + 48 \, (\l \g^{M} \hat{\psi}) \wwb_M \, (\l \g^{m} \hat{\psi}) \, {\partial
\over \partial x^m} \,.
$$

However, for $N\geq 4$, such a gauge is not possible since $\Omega^{(3)}$
and $\Omega^{(4)}$ require non-minimal variables.

Before showing that \brstclo\ is in the BRST cohomology, it will be interesting
to discuss some simple examples of $T$ and the associated vertex operator.
When $N=1$, $V$ becomes independent of $y$ and $\hat\psi$
and describes the antifield of \yind.
When $N=2$, the dilaton
vertex operator $V=i \,\l^{\a j} \l_\a^k \, y_{jk}$ of \purey\ is
obtained from $T = \prod_{J'=1,2} \prod_{\a=1,2} (u_j^{J'} \t^j_\a)$, 
and is dual to the 
super-Yang-Mills action $\int d^4 x  \int du \, \bar D^4 \, {\rm{Tr}}(W^2)$ where
${D'}^4$ and $\bar D^4$ are defined in \nada. 
And when $N=4$, the vertex 
operator 
$V = z^{-2 - \half} (\l \lti)^{-2} 
(\l\g^M\hat\psi)
(\l\g^N\hat\psi)
(\l\g^S\hat\psi) (r \g_{SNM} \lti)$
of footnote 3 is obtained by choosing $T$ to be constant,
and is dual to 
$\int d^4 x \int du \, {D'}^4  \bar D^4 \, {\rm{Tr}} (W^4)$ which is the supersymmetrization 
of the $\int d^4 x \, {\rm{Tr}} (F^4)$ term.

The procedure to show that $V$ of \brstclo\ 
is in the BRST cohomology is as follows: 
To zeroth order in $(\l \g^M \hat{\psi})$  
the condition for $V$ to be annihilated by $Q_\half$ is 
\eqn\first{(z^{\half} \l^{\aba} D_{\aba} + \widetilde{w}^{\aba} r_{\aba})\Omega^{(0)}T = 0 \, .}
To see that $\Omega^{(0)}T$ given in \Omegaum\ satisfies this equation, first note that 
$\l^{\aba} D_{\aba}$  can be decomposed as

\eqn\ldecomposition{ \l^{\aba} D_{\aba} = \wwb_M \www_N  (\l \g^M \g^N D) +  \www_M \wwb_N  (\l \g^M \g^N D) \, .} 
$$ \equiv \l D_1 + \l D_2 $$
\noindent and we can rewrite \first\ as 
\eqn\first{(z^{\half} \l D_2 + \widetilde{w}^{\aba} r_{\aba})\Omega^{(0)}T  + 
z^{\half} [\, \l D_1 \, , \Omega^{(0)} \,] T  = 0 \, }
\noindent where $ [ \, , ] $ means commutator and 
$ (\l D_1) T =0$
since $T$ is G-analytic.
It is easy to see that $[\, \l D_1 \, , \Omega^{(0)} \,] = 0$ since
\eqn\commutator{\{ (\l D_1) , (\lambda \g^N \Dt) \} = 
- 2( \l \g^N \g^m \g^S \l) \wwb_S {\partial
\over \partial x^m} = 0 \, ,} 
\noindent where $\{ \, ,\}$ is an anticommutator and we have used that
$\l$ is a pure spinor. 

To show that $(z^{\half} \l D_2 + \widetilde{w}^{\aba} r_{\aba})\Omega^{(0)}T = 0$ 
we first note that using Fierz identities it is possible to rewrite the first term 
in the right hand side of \Omegaum\ as     
\eqn\Omegaummais{\Omega_{ft}^{(0)} = - {1 \over 4} (\l \g^M \Dt)(\l \g^N \Dt)(\Dt \g_{M N P} \Dt) \www^P} 
$$ - (\l \lti)^{-1} (\l D_2) (\l \g^S \Dt)(\l \g^P \Dt)(\lti \g_{PS} \Dt) $$
\noindent where $ft$ means first term. 
It is easy to see that the second term on 
the right hand side of the equation above is annihilated by $(\l D_2)$. 
To see that the first term is also annihilated, 
it is convenient to choose a Lorentz frame where the only
non-vanishing component of 
$\l$ is $\l^{++}$ which carries $5\over 2$ charge
with respect to a $U(1)$ subgroup of $SO(10)$.
In $SU(5)\times U(1)$ notation, an $SO(10)$ chiral 
spinor $S^\aba$ splits as $(S^{++},S_{ab},S^a)$
where $a=1$ to 5 carrying $U(1)$ charge $({5\over 2}, {\half}, -{3\over 2})$
and an $SO(10)$ vector $v^M$ splits as $(v^a, v_a)$ carrying $U(1)$ charge
$(+1,-1)$.

In this Lorentz frame, the first term of \Omegaummais\ is
\eqn\umm{
C_{1} (\l^{++})^2 (\e_{abcde} \Dt^a \Dt^b \Dt^c \Dt^d v^e) + C_{2}  (\l^{++})^2 (\Dt_{ab} \Dt^a \Dt^b \Dt^c v_c),}

\noindent where $C_{1}$ and $C_{2}$ are constants.
Using that $\l D_2 = \l^{++} (\Dt^a v_a)$ 
and  $v_a v^a=0$,
one verifies that $\l D_2$
annihilates \umm.

 





Also one can prove that       
$$ \widetilde{w}^\aba r_\aba \, \Big{(} \, (\l \lti)^{-2} \, (\l \g^M X)(\l \g^N Y)
(\l \g^P Z) (r \g_{P N M} \lti) \Big{)} = 0 $$

\noindent for any fermionic $X$, $Y$ and $Z$ which implies   
$$ \widetilde{w}^\aba r_\aba \, (\Omega^{(0)}) = - {1 \over 2} (\l \lti)^{-2} (\l D_2) 
(\l \g^M \Dt)(\l \g^N \Dt)(\l \g^P \Dt)(r \g_{P N M} \lti) $$


\noindent and this variation cancels precisely with the action of $z^{\half}(\l D_2)$ in the second
term of the right hand side of $\Omega^{(0)}$  given in \Omegaum\ . This completes the proof that
$(z^{\half} \l D_2 + \widetilde{w}^{\aba} r_{\aba})\Omega^{(0)}T = 0$.       

To see that $V$ is BRST closed in higher orders in $(\l \g^M \hat{\psi})$, one also has to show
that 
\eqn\showne{(N-1)[z^{\half}(\l\g^M\hat{\psi}) \www_M \, \Omega^{(0)} T  + (z^{\half}\l^{\aba} D_{\aba} + 
\widetilde{w}^{\aba} r_{\aba}) \, \Omega^{(1)}  T] = 0,}
$$ (N-1)(N-2)[z^{\half}(\l\g^M\hat{\psi}) \www_M \,  \Omega^{(1)} T 
+ (z^{\half} \l^{\aba} D_{\aba} + \widetilde{w}^{\aba} r_{\aba}) \, \Omega^{(2)} T] = 0, $$ 
$$(N-1)(N-2)(N-3)[z^{\half} (\l\g^M\hat{\psi}) \www_M \, \Omega^{(2)}T + 
(z^{\half} \l^{\aba} D_{\aba} + \widetilde{w}^{\aba} r_{\aba}) 
\, \Omega^{(3)} T ]= 0, $$ 
$$ (N-1)(N-2)(N-3)(N-4) 
[ z^{\half}(\l\g^M\hat{\psi}) \www_M \, \Omega^{(3)}T + 
(z^{\half} \l^{\aba} D_{\aba} + \widetilde{w}^{\aba} r_{\aba}) 
\, \Omega^{(4)}  T] = 0, $$ 
\eqn\finalcondition{ (N-1)(N-2)(N-3)(N-4)(N-5) 
(\l\g^M\hat{\psi}) \www_M \, \Omega^{(4)}T = 0}  
where the 
factors of $ (\l \g^M \hat{\psi}) \www_M$ above come from the BRST variation of
$y_{j k}/z$. 

When $N<4$, only the first two equations of \showne\ need to be satisfied,
and one can show that they are
satisfied by $(\Omega^{(0)}_{min},
\Omega^{(1)}_{min},
\Omega^{(2)}_{min})$ of \omina. However, in the rest of the paper, we will
not put any restrictions on $N$ and will solve all five equations of \showne\
and \finalcondition.

Consider the first equation in \showne. 
To see that $\Omega^{(0)}$ given in  
\Omegaum\ and $\Omega^{(1)}$ given in \omegaone\ satisfy this equation we first
note that \commutator\ implies $ [ \, (\l D_1) \, , \, \Omega^{(1)} ] = 0 $.     
We follow the same steps of the discussion above and rewrite the first term in the right hand side
of \omegaone\ using Fierz identities as



\eqn\omegaonemais{\Omega^{(1)}_{ft} = - (\l \g^M \hat{\psi})(\l \g^S \Dt)(\Dt \g_{M S P} \Dt) \, \www^P}
$$ - (\l \lti)^{-1} (\l \g_M \hat{\psi}) \, \www^M \, (\l \g^S \Dt)(\l \g^P \Dt)(\lti \g_{P S} \Dt) $$
$$ - 3 \, (\l \lti)^{-1} (\l D_2) \, (\l \g^M \hat{\psi})(\l \g^S \Dt)(\lti \g_{S M} \Dt) .  $$

To show that $ (\l \g^M \hat{\psi}) \www_M (\Omega^{(0)}_{ft}) 
+ (\l D_2) (\Omega^{(1)}_{ft}) = 0$, 
note that many terms cancel and one only needs to show that 
\eqn\onlyc{ - {1 \over 4} (\l \g^M \psih) v_M \, (\l \g^M \Dt)(\l \g^N \Dt)(\Dt \g_{M N P} \Dt) \www^P }
$$ - (\l D_2) \, (\l \g^M \psih)(\l \g^S \Dt)(\Dt \g_{M S P} \Dt) \, \www^P  = 0.$$ 
By choosing as before a Lorentz frame in which the only non-vanishing 
component of $\l^{\aba}$ is $\l^{++}$, 
it is not difficult to see that the two terms in \onlyc\ are proportional. 
And we have developed a Mathematica program to fix the coefficients so that
the two terms in \onlyc\ cancel.

We also note that  
\eqn\womegaone{\widetilde{w}^\aba r_\aba \, (\Omega^{(1)}) = -{1 \over 2} \,(\l \lti)^{-2} \, 
(\l \g_M \psih) \, \www^M 
\, (\l \g^N \Dt)(\l \g^P \Dt)(\l \g^S \Dt)(r \g_{S P N} \lti) }
$$ - {3 \over 2} \, (\l \lti)^{-2} \, (\l D_2) \, (\l \g^M \psih)(\l \g^N \Dt)(\l \g^P \Dt)(r \g_{P N M} \lti) $$

\noindent and this variation cancels with the action of $z^{\half}(\l D_2)$ on the $r$ dependent
term of $\O^{(1)}$ and with the action of $z^{\half}(\l \g^M \psih) \www_M$ on the $r$ dependent
term of $\O^{(0)}$. Similar arguments can be given to show that all the equations 
in \showne\ and in \finalcondition\ are satisfied. One last comment is that although 
it may seem surprising that $\Omega^{(4)}$ does not depend on
$r$, this follows because 
\eqn\omegawr{\widetilde{w}^\aba r_\aba \, (\Omega^{(4)}) = -{1 \over 2} (\l \lti)^{-2}
(\l \g_S \psih) \, \www^S \, (\l \g^M \psih)(\l \g^N \psih)(\l \g^P \psih)(r \g_{P N M} \lti)}
\noindent and this variation is precisely cancelled with the action of $z^{\half} 
\www_M (\l \g^M \psih)$ in the $r$
dependent term of $\Omega^{(3)}$, so the last equation of \showne\ is satisfied.   





\subsec{Gauge invariance}

For \brstclo\ to be consistent, $V$ must change by a BRST-trivial quantity under the gauge transformation
of $T$. In other words
$\d V = Q_{1 \over 2} \Sigma$ for some $\S$ when $\d T = 
 (u {\p\over{\p \bar u}})_J^{J'} \Lambda^J_{J'} $.
Integrating \brstclo\ by parts with respect to $D^{J'}_{J}\equiv
(u {\p\over{\p\bar u}})_J^{J'}$ one
finds that 
\eqn\gaugefin{\Sigma = z^{2-{1 \over 2}-N} \int du [(y uu)^{N-1} 
(A^{(0)})_J^{J'} \Lambda^J_{J'} + 8 (N-1)(y uu)^{N-2}
(A^{(1)})_J^{J'} \Lambda^J_{J'}} 
$$ + 8^2(N-1)(N-2)(y uu)^{N-3} 
(A^{(2)})_J^{J'} \Lambda^J_{J'}] $$ 
where 
\eqn\defAo{(A^{(0)})_J^{J'} =3 \, (\l \lti)^{-1} \, (\l \g^M \Dt) 
(\l \g^S \Dt) (\lti \g_{S M}  \Dt^{J'}_{J})} 
$$ + 3 \, (\l \lti)^{-1}  \, \{ \, (\l \g^M  \Dt^{J'}_{J}) \, , (\l \g^N \Dt) \, \}
 (\lti \g_{N M} \Dt) \, ,$$

\eqn\defAum{(A^{(1)})_J^{J'} = 
6 \, (\l \lti)^{-1} \, (\l \g^S \psih) (\l \g^T \Dt) (\lti \g_{T S} \Dt^{J'}_{J})}
$$ + 24 \, (\l \lti)^{-1} \, (\l \g_N \psih) \, \wwb_T \, (\l \, \g^{J'}_{J} \g^m \g^T \g^N \lti) 
{\partial \over \partial x^m} ,$$

\eqn\defAdois{(A^{(2)})_J^{J'} = 3 \, (\l \lti)^{-1} \, 
(\l \g^S \psih)(\l \g^T \psih)(\lti \g_{T S} \Dt^{J'}_{J}),}

\noindent  $ (\Dt^{J'}_{J})_\aba = D^{J'}_{J} (\Dt_\aba) $, and  
$ \g^{J'}_{J} =  D^{J'}_{J} (\wwb_M \, \g^M)$. In order to construct $\Sigma$ 
to satisfy $\d V = Q_{1 \over 2} \Sigma$ we have to solve the following equations  

\eqn\equationsigma{(\l^{\aba} D_{\aba} + 
z^{-\half} \widetilde{w}^{\aba} r_{\aba}) \, (A^{(0)})^{J'}_{J} \, \Lambda^{J}_{J'} 
= (-D_{J}^{J'} \O^{(0)}) \, \Lambda^{J}_{J'} \, ,}
$$ (\l \g^M \psih) \, \www_M \, (A^{(0)})^{J'}_{J} \, \Lambda^{J}_{J'} \, + \,
(\l^{\aba} D_{\aba} + 
z^{-\half} \widetilde{w}^{\aba} r_{\aba}) \,  (A^{(1)})^{J'}_{J} \, \Lambda^{J}_{J'} = 
(-D_{J}^{J'} \O^{(1)} )\, \Lambda^{J}_{J'} \, , $$ 
$$ (\l \g^M \psih) \, \www_M \, (A^{(1)})^{J'}_{J} \, \Lambda^{J}_{J'} \, + \,
(\l^{\aba} D_{\aba} + 
z^{-\half} \widetilde{w}^{\aba} r_{\aba}) \,  (A^{(2)})^{J'}_{J} \, \Lambda^{J}_{J'} = 
(-D_{J}^{J'} \O^{(2)}) \, \Lambda^{J}_{J'} \, , $$  
$$ (\l \g^M \psih) \, \www_M \, (A^{(2)})^{J'}_{J} \, \Lambda^{J}_{J'} 
= 
( -D_{J}^{J'} \O^{(3)}) \, \Lambda^{J}_{J'} \, . $$ 

To see that $(A^{(0)})^{J'}_{J}$ given in \defAo\ satisfies the first equation of \equationsigma, we first 
note that at zero momentum (i.e. setting all the anticommutators to zero) we have
using Fierz identities
\eqn\firstcondition{(-D^{J'}_{J} \Omega^{(0)}) \, \Lambda^{J}_{J'}  =  
3 \, (\l \lti)^{-1} (\l D_2)  \, (\l \g^M \Dt)(\l \g^S \Dt) (\lti \g_{S M} \Dt^{J'}_{J}) \, \Lambda^{J}_{J'} }
$$  - 3 \, z^{-\half} \, (\l \lti)^{-2} \, (\l r) \, (\l \g^N \Dt)(\l \g^P \Dt)(\lti \g_{P N} \Dt^{J'}_{J})  
\, \Lambda^{J}_{J'}  $$
$$ + 3 \, z^{-\half} \, (\l \lti)^{-1} \, (\l \g^M \Dt)(\l \g^P \Dt)(r \g_{P M} \Dt^{J'}_{J}) \, 
\Lambda^{J}_{J'} , $$
 
\noindent where we have used that $\Lambda^{J}_{J'}$ is G-analytic.  
 

Using \firstcondition\ we see that
\eqn\zeromomentum{( \l^{\aba} D_{\aba} + 
z^{-\half} \widetilde{w}^{\aba} r_{\aba}) \, (A^{(0)})^{J'}_{J} \, \Lambda^{J}_{J'}  }
$$ = ( \l D_2  + z^{-\half} \widetilde{w}^{\aba} r_{\aba}) \,  3 \, (\l \lti)^{-1} \, (\l \g^M \Dt) 
(\l \g^S \Dt) (\lti \g_{S M}  \Dt^{J'}_{J})  \, \Lambda^{J}_{J'}  $$
$$ = (-D^{J'}_{J} \Omega^{(0)}) \, \Lambda^{J}_{J'} , $$ 

\noindent so at zero momentum the equation is satisfied. The next step is to
consider the case where the commutators are not zero and we have  
\eqn\nonzerocommutator{ 
(- D^{J'}_{J} \Omega^{(0)})_{ac} \, \Lambda^{J}_{J'} = 3 \,(\l \lti)^{-1} \, (\l D_2) \, 
\{(\l \g^M \tilde{D}^{J'}_{J}) \, ,\, (\l \g^N \tilde{D}) \} (\lti \g_{N M} \Dt) \, \Lambda^{J}_{J'} }
$$ + 3 \, z^{-\half} \, (\l \lti)^{-1} \, \{ (\l \g^M \tilde{D}^{J'}_{J}) \, , \,
(\l \g^N \tilde{D}) \} (r \g_{N M} \tilde{D}) \, \Lambda^{J}_{J'}  $$
$$ - 3 \, z^{-\half} \, (\l \lti)^{-2} \, (\l r) \, \{ (\l \g^M \tilde{D}^{J'}_{J}) \, , \, (\l \g^N \tilde{D}) \}
(\lti \g_{N M} \tilde{D}) \, \Lambda^{J}_{J'} $$
$$ - {3 \over 2} \, (\l \lti)^{-1} \, \{ (\l \g^M \tilde{D}^{J'}_{J}) \, , \, (\l \g^N \tilde{D}) \} \www_N 
(\l \g^S \tilde{D})(\lti \g_{S M} \tilde{D}) \, \Lambda^{J}_{J'} $$
$$ + {3 \over 2} \, (\l \lti)^{-1} \, \www_M \{(\l \g^M \tilde{D}^{J'}_{J}) \, , \, (\l \g^N \tilde{D}) \}
(\l \g^S \tilde{D}) (\lti \g_{S N} \tilde{D})  \, \Lambda^{J}_{J'} $$ 
$$ + {3 \over 2} \, \{ (\l \g^M \tilde{D}^{J'}_{J} ) \, , \, (\l \g^N \tilde{D}) \} 
(\tilde{D} \g_{M N T} \tilde{D}) \www^T  \, \Lambda^{J}_{J'} , $$
\noindent where the subscript $ac$ means the contribution from the
anticommutators.
We note that it is possible to rewrite the 
first three terms of the expression above as 
\eqn\trestermos{3 \,(\l \lti)^{-1} \, (\l D_2) \, 
\{(\l \g^M \tilde{D}^{J'}_{J}) \, ,\, (\l \g^N \tilde{D}) \} (\lti \g_{N M} \Dt) \, \Lambda^{J}_{J'}}
$$+ 3 \, z^{-\half} \, (\l \lti)^{-1} \, \{ (\l \g^M \tilde{D}^{J'}_{J}) \, , \,
(\l \g^N \tilde{D}) \} (r \g_{N M} \tilde{D}) \, \Lambda^{J}_{J'} $$
$$ - 3 \, z^{-\half} (\l \lti)^{-2} \, (\l r) \, \{ (\l \g^M \tilde{D}^{J'}_{J}) \, , \, (\l \g^N \tilde{D}) \}
(\lti \g_{N M} \tilde{D}) \, \Lambda^{J}_{J'} $$
$$ = (\l D_2 + z^{-\half} \bar{w}^{\aba}r_{\aba}) \, 3 \,(\l \lti)^{-1} \,
\{(\l \g^M \tilde{D}^{J'}_{J}) \, ,\, (\l \g^N \tilde{D}) \} (\lti \g_{N M} \Dt) \, \Lambda^{J}_{J'}$$
$$ =  ( \l^{\aba} D_{\aba} + z^{-\half} \bar{w}^{\aba}r_{\aba}) \, 3 \,(\l \lti)^{-1} \,
\{(\l \g^M \tilde{D}^{J'}_{J}) \, ,\, (\l \g^N \tilde{D}) \} (\lti \g_{N M} \Dt) \, \Lambda^{J}_{J'} ,$$

\noindent where we have used 

$$ \{ (\l D_1), \{ (\l \g^M \DJJ), (\l \g^N \tilde{D}) \} 
(\lti \g_{NM} \tilde{D}) \}  = 0$$

\noindent and $(\l D_1) \,  \Lambda^{J}_{J'} = 0$. In order to rewrite the last three
terms of \nonzerocommutator\ in a convenient form we first note

\eqn\principalcommutator{ \{ (\l \g^S \DJJ ) \, , \, (\l \g^P \Dt) \} = 
\, 2 \, \wwb_T \, (\l \g^S \g^{J'}_{J} \g^m \g^T \g^P \l) \, {\partial \over \partial x^m}}

\noindent and this implies   

\eqn\identi{\{ (\l \g^M \tilde{D}^{J'}_{J}) \, , \, (\l D_2) \} \,  = -
 \{ (\l \g^N \tilde{D}^{J'}_{J}) \, , \, (\l \g^M \tilde{D}) \} \, \www_N .} 



Using the identity of \commutator\ we have  

$$ \{ (\l \g^N \tilde{D}^{J'}_{J}) \, , \, (\l D) \}  = 0 $$

\noindent and so

\eqn\identidois{ \{ (\l \g^S \tilde{D}^{J'}_{J}) \, , \, (\l D_2) \} 
= - \{ (\l \g^S \tilde{D}^{J'}_{J} ) \, , \, (\l D_1) \} .}

Also noting that 

$$ 3 \, (\l \lti)^{-1} \, (\l \g^M \DJJ) (\l \g^S \tilde{D}) \{ (\l D_1) \, , \,
(\lti \g_{S M} \tilde{D}) \} $$ 
$$ + 6 \{ (\l D_1) \, , \, 
(\l \g^S \tilde{D})(\tilde{D} \g_S \DJJ ) \} $$
$$ = {3 \over 2} \{ (\l \g^M \DJJ) \, , \, (\l \g^N \tilde{D}) \} (\tilde{D} \g_{M N T} \tilde{D} ) \www^T ,$$  
we see that 
\eqn\moreidentity{ - {3 \over 2} \, (\l \lti)^{-1} \, \{ (\l \g^M \tilde{D}^{J'}_{J}) \, , \, (\l \g^N \tilde{D}) \} \www_N 
(\l \g^S \tilde{D})(\lti \g_{S M} \tilde{D}) \, \Lambda^{J}_{J'} }
$$ + {3 \over 2} \, (\l \lti)^{-1} \, \www_M \{(\l \g^M \tilde{D}^{J'}_{J}) \, , \, (\l \g^N \tilde{D}) \}
(\l \g^S \tilde{D}) (\lti \g_{S N} \tilde{D})  \, \Lambda^{J}_{J'} $$ 
$$ + {3 \over 2} \, \{ (\l \g^M \tilde{D}^{J'}_{J} ) \, , \, (\l \g^N \tilde{D}) \} 
(\tilde{D} \g_{M N T} \tilde{D}) \www^T  \, \Lambda^{J}_{J'} $$
$$ = \{ (\l D_1) \, , \, (A^{(0)})^{J'}_{J} \} ,$$ 

\noindent where we used the fact that the first part $fp$ of $(A^{(0)})^{J'}_{J}$ 
(i.e. the part that is independent of the commutators)
can be written in the form 
given below using Fierz identities 
$$
(A^{(0)}_{fp})^{J'}_{J} = 3 \, (\l \lti)^{-1} \, (\l \g^M \DJJ)(\l \g^S \tilde{D})(\lti \g_{S M} \tilde{D}) $$
$$ + 6 (\l \g^M \tilde{D})(\tilde{D} \g_M \DJJ ). $$
Using \zeromomentum, \trestermos\ and \moreidentity\ we see that
$$(\l^{\aba} D_{\aba} + 
z^{-\half} \widetilde{w}^{\aba} r_{\aba}) \, (A^{(0)})^{J'}_{J} \, \Lambda^{J}_{J'} 
= (-D_{J}^{J'} \O^{(0)}) \, \Lambda^{J}_{J'} .$$
Similar arguments can be used
to show that all the equations given in \equationsigma\
are satisfied.

\newsec{Summary and Possible Applications}

In this paper, we expanded the BRST operator near the boundary of $AdS_5$
and explicitly computed the zero mode cohomology corresponding to
supergravity states. The leading behavior near the $AdS_5$ boundary of
supergravity vertex operators was expressed in ${\cal N}=4$ $d=4$
harmonic superspace in \brstclo\
and was shown to be BRST-closed and gauge invariant.

There are several possible applications of our results. One possible
application is to generalize our methods to massive $AdS_5\times S^5$
vertex operators and to compute the spectrum. This generalization would
require allowing dependence on nonzero modes of the worldsheet variables
in the analysis of the BRST cohomology. It would be interesting to compare
the resulting vertex operator for the Konishi state with 
the pure spinor and RNS vertex operators proposed in \vallilo\ and \minahan.

Another possible application is
to use the supergravity vertex operators to compute tree-level
superstring scattering amplitudes corresponding to planar super-Yang-Mills
correlation functions. For generic tree amplitudes, these computations
would probably require working out the behavior of the supergravity
vertex operators away from the boundary of $AdS_5$.
However, knowing the boundary behavior may be enough for computing certain
terms in disc
amplitudes with one closed string supergravity vertex operator in the bulk
and $N$
open string super-Yang-Mills vertex operators located on D-branes near
the $AdS_5$ boundary. For the supergravity vertex operator constructed from
$T^{(4-N)}$ in \brstclo,
one expects the resulting disc amplitude to contain terms proportional to
$\int d^4 x \int du \int d^8 (u\t) W^{(N)}(u,x,\t)
T^{(4-N)}(u,\bar u,x,\t)$ of \harmint.

\vskip15pt
{\bf Acknowledgements:}
We would
like to thank Thales Agricola, Renann Jusinskas, Juan Maldacena, Luca 
Mazzucato, Andrei Mikhailov and Brenno Vallilo
for useful discussions, NB would like to thank
CNPq grant 300256/94-9
and FAPESP grants 2009/50639-2 and 2011/11973-4 for partial financial support,
and TF would like to thank 
FAPESP grant 2009/50775-3 for financial support.

\listrefs
\end